# Bifurcations of self-similar solutions of the Fokker-Plank Equation

F. Berezovskaya[1], G. Karev[2]

[1] Department of Mathematics, Howard University, Washington D.C., 20059.

[2] National Center for Biotechnology Information, National Library of Medicine, National Institutes of Health, Bethesda, MD 20894

Email: fsberezo@hotmail.com, karev@ncbi.nlm.nih.gov

**Abstract.** A class of one-dimensional Fokker-Plank equations having a common stationary solution, which is a power function of the state of the process, was found. We prove that these equations also have generalized self-similar solutions which describe the temporary transition from one stationary state to another. The study was motivated by problems arising in mathematical modeling of genome size evolution.

## 1. Motivations and statement of the problem

A broad variety of phenomena in physics, biology, economics, etc. is described by power law distributions. Recent studies have shown that the distributions of many genome-related quantities could be well described by the so-called Pareto distribution: $P(i) = c(i+a)^{-\gamma}$ where $\gamma>0$, $a$ are parameters (Gisiger et al, 2001; Qian et al, 2001; Koonin et al, 2002). In our previous work (Karev et al., 2002, 2003), we have shown that well-known birth-and-death processes are a natural source of the power solutions; a special class of such processes (models BDIM, after <u>b</u>irth, <u>d</u>eath and <u>i</u>nnovation <u>m</u>odel) shows the power law distribution of its stationary solutions, which are consistent with known data on the sizes of gene families. The analysis of stochastic BDIMs (Karev et al., 2004) showed that non-linear versions of such models can well approximate not only the size distribution of gene families but also the dynamics of their formation during genome evolution. The fact that only higher degree BDIMs are compatible with the observed characteristics of genome evolution suggests that the growth of gene families is self-accelerating, which might reflect differential selective pressure acting on different genes.

However, even non-linear BDIMs give unreasonable estimation of the average time of formation of the largest gene families ($10^{11}$ yrs compared to the realistic $\sim 10^9$) and only the minimal time (about $2.5 \times 10^9$ yrs) is close to reality. Thus, the problem arises: can we find a different, not purely stochastic approach to model rapid increase of genome size? To examine this problem, we formulate a diffusion approximation of the BDIM (it is well known that a birth-and-death process with discrete space of states is an analog of a diffusion process with a continuous space of states, and vice versa).

In the framework of the diffusion model, we found generalized self-similar solutions (gss) which could be interpreted as the process of deterministic self-accelerating increase of the genome size.

This paper is organized as follows. In section 2, we describe the Fokker-Plank equation (FPE) corresponding to the non-linear BDIM; the stationary solutions of this FPE are considered in section 3. An important example of the linear diffusion model is considered in section 4. In section 5, we describe a class of non-linear FPEs which have a given common stationary solution. The main section 6 contains the definition and investigation of the generalized self-similar solutions of the FPE. Section 7 contains a brief discussion of obtained solutions. The proofs of main theorems are given in the Appendix.

## 2. Diffusion version of the BDIM and the Fokker-Plank equation

Let a population be subdivided into $N$ (finite or infinite) different groups, which we will call "families" and $f(x,t)$ be the number of families of the size $x$ in $t$ time moment. Let us suppose that the "individual" birth and death rates in a family of the size $x$ are $\lambda(x)$ and $\delta(x)$ respectively. Then the equation

$$\frac{\partial f(t,x)}{\partial t} = f(t,x-1)\lambda(x-1) - (\lambda(x)+ \delta(x))f(t, x)+ \delta(x+1)) f(t, x+1)], \quad x>0, \qquad (2.1)$$

(subject to boundary conditions) describes the birth-and-death process with the set of states $\{0,1,\ldots N\}$.

A formal continuous approximation of equation (2.1) gives the equation:

$$\frac{\partial f(t,x)}{\partial t} = -\frac{\partial}{\partial x}[f(t, x) \mu(x)] + \frac{1}{2}\frac{\partial^2}{\partial x^2}[f(t, x)\sigma^2(x)] \qquad (2.2)$$

where $\mu(x)=\lambda(x)-\delta(x)$ is the drift and $\sigma^2(x)=\lambda(x)+\delta(x)$ is the diffusion coefficient. Equation (2.2) is the *Fokker-Plank equation* (FPE) for the considered process.

**Remark.** The problem of "equivalence" of models (2.1) and (2.2) may be non-trivial (see, e.g., Gardiner, 1985) and we do not discuss it here.

Let us denote $J(x,t)$ the *current* of particles through the point $x$ at time moment $t$

$$J(t, x) = f(t, x) \mu(x) - \frac{\partial}{\partial x}[\frac{1}{2} f(t,x) \sigma^2(x)]. \qquad (2.3)$$

Then the Fokker-Plank equation (2.2) could be written as the equation of continuity (or the equation of mass conservation):

$$\frac{\partial f(t,x)}{\partial t} = -\frac{\partial}{\partial x} J(t, x), \qquad (2.4)$$

To solve this equation, we need an initial condition and boundary conditions at the ends of the interval $[r, N]$. For example, if the system is "closed", i.e., a particle cannot leave the interval and there is zero net flow across the ends, then $J(t, x)=0$ at $x=r$ and $x=N$. (2.5)

If the system is "open" and innovation is possible, e.g., at the left end, then the current $J(t, r)$ through the left end, or the rate of innovation $v(t) = -\frac{\partial}{\partial x}f(t,x)\lambda(x)|_{x=r}$ could be taken as the boundary condition of Fokker-Plank equation (2.2) or (2.4).

Typically, the diffusion coefficient $\sigma^2(x)$ is smaller or, at most, of the same order as the drift $\mu(x)$. For example, as follows from formula (3.1) below, if $\sigma^2(x)=$const and $\mu(x)=cx$, $c<0$ is a constant, then its stationary solution follows the (truncated) normal distribution; if $\sigma^2(x)=$const and $\mu(x)=c<0$, then its stationary solution follows the (truncated) exponential distribution.

In this paper, we explore the diffusion approximation (2.2) of the polynomial and rational BDIMs which have been previously considered for a discrete phase space (Karev et al., 2002, 2003). We show that the stationary solution $f_{st}(x)$ of model (2.2) follows the Pareto distribution only when $\sigma^2(x)$ increases faster then $|\mu(x)|$ and the drift is negative as opposed to the usually considered case. Next, the problem of estimation of the duplication/deletion rates of genes is hard; hence the birth and death rates or drift and diffusion coefficients in corresponding models of genome size evolution are actually unknown. By contrast, the distributions of gene family sizes are well-established empirical data. Thus, we constructed and explored a class of diffusion models that have a given common stationary solution but different drift and diffusion coefficients. We proved that these models have special "self-similar" solutions that describe the transitions from one stationary solution to another. A speed of the movement of the "front" of this solution depends on the model parameters.

### 3. Stationary solution of the model and the power asymptotics

The stationary solution $f_{st}(x)$ of model (2.2), for which $df_{st}(x)/dt=0$, satisfies the equation

$$-\frac{\partial}{\partial x}[f_{st}(x)\mu(x)] + \frac{1}{2}\frac{\partial^2}{\partial x^2}[f_{st}(x)\sigma^2(x)] = dJ(x)/dx = 0,$$

so the current $J(x)$=const at all $x$.
If the system is closed and hence $J(x)=0$ at $x=r$ (due to the boundary condition (2.5)) then

$$f_{st}(x)\mu(x) = \frac{1}{2}\frac{\partial}{\partial x}[f_{st}(x)\sigma^2(x)] \text{ for all } x \in [r,N].$$

The solution of this equation is

$$f_{st}(x) = \frac{\sigma^2(r)f_{st}(r)}{\sigma^2(x)} \exp\left(2\int_r^x \frac{\mu(u)}{\sigma^2(u)}du\right). \qquad (3.1)$$

If the system is open and $J(x)\neq 0$ at $x=r$, then other stationary solutions can exist (see, e.g., Gardiner, ch.5). The following assertion easily follows from (3.1).

**Theorem 1**. Let $[\mu(x)+1/2\frac{\partial}{\partial x}\sigma^2(x)]/\sigma^2(x) = -\gamma/(2x)+S(x)$, where $\gamma$ is a constant and $S(x)$ satisfies the following condition: $\int_r^x S(y)dy \to const$ at $x\to\infty$. Then $f_{st}(x) \sim x^{-\gamma}$.

**Corollary 1.** Let $\sigma^2(x) = x^\rho(a+o(1/x))$, $\rho$, $a>0$ are constants; let $\mu(x)/\sigma^2(x) = -\eta/(2x)+O(1/x^2)$. Then $f_{st}(x) \sim x^{-\eta-\rho}$.

As a representative example, let us consider a *linear* diffusion model with $\lambda(x)$ and $\delta(x)$ being linear functions of $x$: $\lambda(x)=\lambda(x+a)$, $\delta(x)=\delta(x+b)$ where $\lambda$ and $\delta$ are positive constants. Then $\mu(x) = (\lambda - \delta)x + \lambda a - \delta b$, $\sigma^2(x) = (\lambda + \delta)x + \lambda a + \delta b$.
Suppose also that the constants $a$, $b$ are such that $\sigma^2(x) > 0$ in $[r, N]$.

The linear discrete-state BDIM (Karev et al., 2002) has a stable distribution $f_{st}(x)$, which is asymptotically equal to the power-law distribution if and only if $\lambda = \delta$; then $f(x) \sim x^{-\gamma}$, where $\gamma = b-a+1$. Similar result is valid also for diffusion linear model

$$\frac{\partial f(x,t)}{\partial t} = \{-\frac{\partial}{\partial x}[f(x,t)((\lambda-\delta)x+\lambda a-\delta b)] + \frac{1}{2}\frac{\partial^2}{\partial x^2}[f(x,t)((\lambda+\delta)(x+s))]\} \quad (3.2)$$

where $s=(a\lambda+b\delta)/(\lambda+\delta)$.

**Proposition 1**. *For linear diffusion model (3.2), $f_{st}(x) = C \exp(-lx)(s+x)^{-\gamma}$*

where $\gamma = 4(b-a)\dfrac{\lambda\delta}{(\lambda+\delta)^2} + 1$, $C = f_{st}(r)\exp(lr)(s+r)^{\gamma+1}$, and $l = 2(\delta-\lambda)/(\lambda+\delta)$.

**Corollary 2.** *The stationary distribution of the linear diffusion model is a Pareto distribution if and only if $\lambda = \delta$ and $a-b \neq 1$; under these conditions, $f_{st}(x) = c_1(s+x)^{-\gamma}$ where $\gamma = b-a+1$, $s=(a+b)/2$ and $c_1 = f_{st}(r)(s+r)^{\gamma}$.*

Next, let us consider a more general case and suppose that birth and death rates are polynomials on $x$. Informally, polynomial models could take into account interactions between particles and reflect a feedback between the family size and growth rate (these models with discrete space of states was studied in (Karev et al., 2001, 2003)). Formula (3.1) yields an explicit stationary distribution for polynomial models because the rational function $\mu(x)/\sigma^2(x)$ is integrable.
The problem of the most interest is the asymptotical behavior of the stationary distribution. This behavior critically depends on the relation between the degrees of the polynomials.

**Proposition 2**. *Let the birth and death rates be of the form $\lambda(x) = \lambda x^\rho R(x)$, $\delta(x) = \delta x^\rho Q(x)$, where $R(x)$, $Q(x)$ are polynomials of the same degree $m \geq 1$,*

$$R(x) = \sum_{s=0}^{m} r_s x^{m-s}, \quad Q(x) = \sum_{s=0}^{m} q_s x^{m-s}, \quad r_0 = q_0 = 1 \text{ and } \rho > 0.$$

*Then $f_{st}(x) \sim \exp(-lx) x^{-\nu-m}$,*
*where $l = 2(\delta-\lambda)/(\delta+\lambda)$, $\nu = 4(q_1-r_1)$ and $f_{st}(x) \sim x^{-\nu-m}$, if $\delta = \lambda$.*

4. **Spatial-temporal dynamics of the linear model; a special class of solutions**

Let us consider the linear model (3.2) with $\lambda = \delta$; in this case $\mu(x) = \lambda(a-b)$, $\sigma^2(x) = \lambda(2x+a+b)$. After simple algebra, equation (3.2) reads

$$\frac{\partial f(x,t)}{\lambda \partial t} = (1+\gamma)\frac{\partial f}{\partial x} + (x+s)\frac{\partial^2 f}{\partial x^2}. \quad (4.1)$$

where $\gamma = b-a+1$, $s=(a+b)/2$.
The stationary solution of this model is
$$f_{st}(x) = c_1(s+x)^{-\gamma} \quad (4.2)$$
where $c_1 = f_{st}(r)(s+r)^{\gamma}$; its current is

$$J(t,x) = -\lambda \gamma f(t,x) - \lambda(x+s) \frac{\partial}{\partial x} f(t,x).$$

The following theorem describes a special case of "self-similar" solutions.

**Theorem 2.** *Linear model* (4.1) *with* $N=\infty$ *for any* $t_0 \geq 0$ *has a solution*

$$f(x,t) = (x+s)^{-\gamma} [C_1 + C_2 \Gamma(\gamma, \frac{x+s}{\lambda t})] \tag{4.3}$$

*where* $\gamma = b-a+1$, $s=(a+b)/2$ *and* $\Gamma(a,x) = \int_x^\infty \exp(-t) t^{a-1} dt$ *is the incomplete Gamma-function.*

Formula (4.3) describes the transformation of the initial stationary solution $f_{st}(x) = C_1(x+s)^{-\gamma}$ into another stationary solution of the same "Pareto shape", $f_{st}(x) = (C_1 + C_2 \Gamma(\gamma))(x+s)^{-\gamma}$. Indeed, according to the properties of incomplete $\Gamma$-function, $\Gamma(\gamma, 0) = \Gamma(\gamma)$, $\Gamma(\gamma, \infty) = 0$. Hence, $f(x,0) = C_1(x+s)^{-\gamma}$, and $f(x,t) \to (C_1 + C_2 \Gamma(\gamma))(x+s)^{-\gamma}$ at $t \to \infty$.

More general results are proved below (see Theorems 3 and 4).

## 5. Transformations of the model

The initial point of our investigation was the stationary distribution of sizes of domain families, which was extracted from the empirical data and followed the (truncated) Pareto distribution. Here we construct a class of models with different dynamics but with a common, fixed stationary distribution.

The diffusion model

$$\frac{\partial f(t,x)}{\partial t} = -\frac{\partial}{\partial x}[f(t,x)(\lambda(x)-\delta(x))] + \frac{1}{2}\frac{\partial^2}{\partial x^2}[f(t,x)(\lambda(x)+\delta(x))] \tag{5.1}$$

has the stationary solution (3.1) with $\mu(x) = \lambda(x) - \delta(x)$, $\sigma^2(x) = \lambda(x) + \delta(x)$, $r < x < N$.

Let $g(x)$ be a positive smooth function; transform the initial birth and death rates $\lambda(x)$ and $\delta(x)$ using formulas

$$\lambda^*(x) = \lambda(x)g(x) + (\delta(x) + \lambda(x))/4 \, \partial g(x)/\partial x = \lambda(x)g(x) + \sigma^2(x)/4 \, \partial g(x)/\partial x, \tag{5.2}$$
$$\delta^*(x) = \delta(x)g(x) - (\delta(x) + \lambda(x))/4 \, \partial g(x)/\partial x = \delta(x)g(x) - \sigma^2(x)/4 \, \partial g(x)/\partial x.$$

We suppose that the function $g(x)$ is such that $\lambda^*(x)$ and $\delta^*(x)$ are non-negative for $r < x < N$.

**Proposition 3.** *The stationary solutions of the initial diffusion model* (5.1) *and transformed model with*

$$\sigma^{*2}(x) = \lambda^*(x) + \delta^*(x) = \sigma^2(x) g(x), \tag{5.3}$$
$$\mu^*(x) = \mu(x)g(x) + \sigma^2(x)/2 \, \partial g(x)/\partial x$$

*are identical up to the normalizing constant.*

The proof easily follows from formula (3.1).

For the given functions $\mu(x)$ and $\sigma^2(x)$, let us define the operator

$$J[f](t,x) = f(t,x)\mu(x) - \frac{1}{2}\frac{\partial}{\partial x}[f(t,x)\sigma^2(x)].$$

If the function $f(t,x)$ satisfies the FPE (2.4), then $J[f]$ is the current for diffusion model (2.2). Denote $J^*[f]$ the operator corresponding to the transformed functions $\mu^*(x)$ and $\sigma^{*2}(x)$.

**Lemma 2**. $J^*[f](t,x)=g(x) J[f](t,x)$.

Let us explore the transformations of the linear diffusion model (4.1) which has the stationary solution (4.2). The current for this model is $J[f](t,x)=$
$-\lambda[\gamma f(t,x)+(x+s)\frac{\partial}{\partial x}f(t,x)])$, hence the current for the transformed model is

$$J^*[f](t,x) = -\lambda g(x) [\gamma f(t,x)+ (x+s)\frac{\partial}{\partial x} f(t,x)].$$

Thus, the FPE for the transformed model is $\frac{\partial}{\partial t}f(t,x)=-\frac{\partial}{\partial x} J^*[f](t,x)$, or

$$\frac{\partial}{\lambda\partial t}f(t,x)= \frac{\partial}{\partial x} g(x)[\gamma f(t,x)+ (x+s)\frac{\partial}{\partial x} f(t,x)]. \qquad (5.4)$$

Below we explore the transformation of the linear diffusion model using the function $g(x)=(x+s)^{\rho-1}$, $\rho \geq 1$, $b>(\rho-1)/2$. Then
$\lambda^*(x)= (x+a+ (\rho-1)/2) (x+s)^{\rho-1}$, $\delta^*(x)= (x+b-(\rho-1)/2) (x+s)^{\rho-1}$, and
$\mu^*(x) = (\rho-\gamma)(x+s)^{\rho-1}$, $\sigma^{*2}(x) =2(x+s)^{\rho}$. The corresponding FPE is of the form

$$\frac{\partial}{\lambda\partial t}f(t,z)= z^{\rho} \{\frac{\partial^2}{\partial z^2} f(t,z)+ (\gamma+\rho)/z\frac{\partial}{\partial z} f(t,z) + [\gamma(\rho-1)/z^2]f(t,z)\} \qquad (5.5)$$

where $z=x+s$. Equation (5.5) has a set of stationary solutions:

$f_{st}(t,z)=$
$z^{-(\gamma+\rho-1)/2}\{C_1 BesselJ[(\gamma+\rho-1)/2,(\gamma(\rho-1))^{1/2}]+C_2 BesselY[(\gamma+\rho-1)/2,(\gamma(\rho-1))^{1/2}]\}$.

## 6. Spatial-temporal dynamics of the transformed models

A solution $f(x,t)$ of equation (2.2) is called *generalized self-similar* (gss) if it is of the form $f(x,t) = x^a G(y)$ where $y =x/\phi(t)$ with smooth $\phi(t)\neq 0$ and $a$ is a (real) constant.

**Theorem 3.** *Let $C_1$, $C_2$ be arbitrary constants and $t_0 \geq 0$.*

i) *For $\rho<2$, equation (5.5) has a three-parametric $(C_1,C_2, t_0)$-family of gss-solutions*
$$f(t,x)=(x+s)^{-\gamma} [C_1 + C_2\Gamma(1+(\gamma-1)/(2-\rho),1/(2-\rho)^2 (x+s)^{2-\rho}/(\lambda(t+ t_0)))]. \qquad (6.1)$$

ii) *For $\rho=2$, equation (5.5) has a four-parametric $(C_1, C_2, t_0, \alpha)$-family of gss-solutions*
$$f(x,t)=(x+s)^{-\gamma}\{C_1+ C_2 [(x+s) \exp(-\alpha\lambda(t+ t_0))]^{\gamma-\alpha-1}\} \qquad (6.2)$$
*where $\alpha$ is an arbitrary constant such that $\alpha \neq \gamma-1$, and three-parametric $(C_1, C_2, t_0)$-family of gss-solution,*
$$f(x,t) = (x+s)^{-\gamma} ( C_1+ C_2 (\log(x+s) - (\gamma-1)\lambda(t+ t_0))), \qquad (6.3)$$
*if $\alpha = \gamma-1$.*

iii) *For $2<\rho<1+\gamma$, equation (5.5) has $(C_1, C_2, t_0)$-family of gss-solutions*
$f(t,x)=$ $\qquad (6.4)$
$(x+s)^{-\gamma}\{C_1+C_2((\lambda(t+t_0))^{1/(\rho-2)}(x+s))^{\gamma-\rho+1}E[(\gamma-1)/(\rho-2),1/(\lambda(t+t_0)(s+x)^{\rho-2}(\rho-2)^2)]\}$

*where $E[u,v]= \int_1^\infty \exp(-vt)t^{-u} dt$ is a special function, the Exponential Integral.*

iv) *For $\rho > 1+\gamma$, equation (5.5) has $(C_1, C_2, t_0)$-family of gss-solutions*
$$f(t,x) = (x+s)^{-\gamma} [C_1 + C_2\Gamma(1-(\gamma-1)/(\rho-2), 1/((\rho-2)^2 (x+s)^{\rho-2}\lambda(t+t_0)))]. \quad (6.5)$$

**Theorem 4.** *Let $t_0=0$; then for any fixed value of x,*
1. *gss-solutions (6.1) and (6.5) are bounded functions of t at $t \to 0$, $t \to \infty$ for any constants $C_1$, $C_2$;*
2. *gss-solution (6.2) is a bounded function of t at $t \to 0$, $t \to \infty$ for any value of $\alpha$ such that $0 < \alpha < \gamma-1$ or $0 > \alpha > \gamma-1$ and any constants $C_1$, $C_2$;*
<u>*if the constant $C_2 \neq 0$, then any solution (6.2) with $\alpha < \gamma-1 < 0$ or with $\alpha > \gamma-1 > 0$ as well as any solution (6.3) with $\gamma \neq 1$ is unbounded at $t \to \infty$ ;*</u>
3. *For any constants $C_1$ and $C_2 \neq 0$, gss-solution (6.4) is unbounded at $t \to \infty$.*

Proofs of Theorems 3 and 4 are given in the Appendix.

## 7. Discussion

Asymptotic behavior of generalized self-similar solutions of FPE (5.1), in which diffusion and drift coefficients are rational functions $\sigma^2(x) = x^\rho(a+O(1/x))$ where $a>0$, and $\mu(x) = x^{\rho-1}(b+O(1/x))$, respectively, is essentially determined by the "degree of non-linearity" $\rho \geq 1$ and the first coefficients $a$, $b$ of their expansions. If $\mu(x)/\sigma^2(x) = -2\gamma/x + O(1/x^2)$ where $\gamma > 0$, the FPE has a power stationary solution $f_{st}(x) \sim x^{-\gamma}$. Given the stationary solution, we can construct a class of FPE which have this stationary solution. For this class of FPE, we found families of generalized self-similar solutions $f(t,x)$ of (5.1). Let us emphasize that gss-solutions describe the regime of spreading of the "profile", or "front" of the solution along the $x$ axis with time (in terms of the original problem, this solution describes the increase of genome size). The speed and direction of the movement of these solutions can be defined from the relation $f(t,x)$=const.

In the context of modeling genome size evolution, we are interested mainly in bounded solutions of the model. In particular, solution (6.1) ($1 \leq \rho < 2$) describes the transition from the initial stationary distribution $f(0,x) = C_1(x+s)^{-\gamma}$ to the final stationary distribution $f(\infty,x) = (x+s)^{-\gamma}[C_1 + C_2\Gamma(1+(\gamma-1)/(2-\rho))]$ which differs from the initial one only by a constant multiplier. The front of solution (6.1) moves to the right if $C_2 > 0$.

In the case $\rho=2$ the behavior of gss-solutions can substantially change. Any solution (6.2) is bounded only if $0 \leq \alpha < \gamma+1$. In this case, the solution describes the transition from the initial stationary distribution $f(0,x) = C_1\underline{(x+s)^{-\gamma}} + C_2\underline{(x+s)^{-\alpha-1}}$ to the final stationary distribution $f(\infty,x) = C_1\underline{(x+s)^{-\gamma}}$. Let us note that the initial and final solutions now have different shapes; the front of this solution moves to the right if $C_2 > 0$.

For $2 < \rho < 1+\gamma$, all gss-solutions (6.4) (which are different from the stationary one) tend to infinity in any point $x$ at $t \to \infty$.

Finally, if $\rho > 1+\gamma$, gss-solution (6.5) describes the transition from the initial stationary distribution $f(0,x)=C_1(x+s)^{-\gamma}$ to the final stationary distribution $f(\infty,x)= (x+s)^{-\gamma}[C_1 + C_2\Gamma(1-(\gamma-1)/(\rho-2))]$, similarly to the solution (6.1).

Let us note that the "structure" of the self-similar variable $y = x/t^\beta$ abruptly changes at $\rho=2$, namely, the exponent $\beta>0$ for $1 \leq \rho < 2$ and $\beta < 0$ for $\rho > 2$. We suspect that the case $\rho \leq 2$ is more suitable for modeling genome size evolution.

## Appendix

**Proof of Theorem 3.**
i) Let $\rho \geq 1$, $\rho \neq 2$. Searching a gss-solution of equation (5.5) in the form:
$$f(z, \tau) = z^a G(y) = z^a G(z\tau^{-\beta}) \qquad (A.1)$$
where $z = x+s$, $\tau = \lambda(t+t_0)$ and constants $a$, $\beta$ should be determined, one easily gets that $G(y)$ must satisfy the equation
$$y^2 G_{yy}(y) + ((y(2a+\gamma+\rho)+\beta y^\kappa) G_y(y) + (a+\rho-1)(a+\gamma)G(y) = 0, \quad (A.2)$$
for $\beta = 1/(2-\rho)$ and $\kappa = 3-\rho$. Choosing $a = -\gamma$ we get
$$y G_{yy}(y) + (((\rho-\gamma) + y^{2-\rho}/(2-\rho)) G_y(y) = 0 \qquad (A.3)$$
Equation (A.3) has the following general solution:
$$G(y) = C_1 + C_2 y^{\gamma-\rho+1} \int_1^\infty t^{-u} e^{-vt} dt = C_1 + C_2 y^{1+\gamma-\rho} E[u,v], \qquad (A.4)$$
where $v = y^{2-\rho}/(2-\rho)^2$ and $u = (\gamma-1)/(\rho-2)$.
Thus, equation (5.5) has the gss-solution $f(t,x) =$
$(x+s)^{-\gamma} \{C_1 + C_2((\lambda(t+t_0))^{1/(\rho-2)}(x+s))^{\gamma-\rho+1} E[(\gamma-1)/(\rho-2),(s+x)^{2-\rho}/(\lambda(t+t_0)(\rho-2)^2)]\}$.
For $\rho < 2$ and $\rho > 1+\gamma$ (in both cases $1-u > 0$), solution (A.4) can be written in the form
$$G(y) = C_1 + C_3 \Gamma[1-u, v] = C_1 + C_3 \Gamma[1+(\gamma-1)/(2-\rho), y^{2-\rho}/(2-\rho)^2] \qquad (A.5)$$
and corresponding gss-solution read
$$f(t,x) = (x+s)^{-\gamma} [C_1 + C_2 \Gamma(1+(\gamma-1)/(2-\rho), 1/(2-\rho)^2 (x+s)^{2-\rho}/(\lambda(t+t_0)))].$$
2) Let $\rho = 2$; searching a gss-solution of equation (5.5) in the form:
$$f(z,t) = z^{-\gamma} G(y) \text{ where } y = z \exp(-\alpha\tau) \qquad (A.6)$$
one can show that $G(y)$ must satisfy the equation
$$y G_{yy}(y) + (2-\gamma+\alpha) G_y(y) = 0.$$
This equation has a general solution
$$G(y) = C_1 + C_2 y^{-(\alpha+1-\gamma)} \qquad (A.7a)$$
if $\alpha+1-\gamma \neq 0$ and
$$G(y) = C_1 + C_2 \ln y \qquad (A.7b)$$
if $\alpha = \gamma-1$.
Thus, equation (5.5) has the gss-solution
$$f(t,x) = C_1 (x+s)^{-\gamma} + C_2 (x+s)^{-\alpha-1} \exp(\alpha(\alpha-\gamma+1)\lambda(t+t_0))) \text{ if } \alpha+1-\gamma \neq 0 \text{ and}$$
$$f(t,x) = (x+s)^{-\gamma-1} [C_1 + C_2(\ln(x+s) - \alpha\lambda(t+t_0))] \text{ for } \alpha+1 = \gamma.$$
Q.E.D.

**Proof of Theorem 4.**
Recall that $\Gamma(u,v) \to 0$ at $v \to \infty$ and $\Gamma(u,v) \to \Gamma(u)$ at $v \to 0$ for positive $u$. Hence, for $\rho < 2$ and for $\rho > 1+\gamma$, $\Gamma(1+(\gamma-1)/(2-\rho), 1/(2-\rho)^2 (x+s)^{2-\rho}/(\lambda t)) \to 0$ at $t \to 0$ and at $t \to \infty$ $\Gamma(1+(\gamma-1)/(2-\rho), 1/(2-\rho)^2 z^{2-\rho}/(\lambda t)) \to \Gamma(1+(\gamma-1)/(2-\rho)) > 0$. Hence, $f(0,x) = C_1(x+s)^{-\gamma}$ and $f(t,x)$ tends to $f(\infty,x) = (x+s)^{-\gamma}[C_1 + C_2\Gamma(1+(\gamma-1)/(2-\rho))]$ at $t \to \infty$. So, solution (6.1) describes the transition from the initial stationary distribution $f(0,x)$ to the final stationary distribution $f(\infty,x)$, which differs from the initial one only by a constant multiplier.
For $\rho = 2$, gss-solutions depend on a free parameter $\alpha$. Solution (6.2) is bounded for all $t$ only if $0 \leq \alpha < \gamma-1$. It is easy to see that, at $\alpha \to \gamma-1$, the solution (6.2) transforms to (6.3) which is unbounded at $t \to \infty$.
It is known that $e^v E[u,v] = O(1/v)$ (Abramovitz, Stegun, 1970, 5.1.19), hence $v^s E[u,v] \to 0$ at $v \to \infty$ for any $s > 0$. When this relation is applied to solution (6.4) for $2<\rho<1+\gamma$, the term $C_2((\lambda t)^{1/(\rho-2)}z)^{\gamma-\rho+1}E[(\gamma-1)/(\rho-2),1/(\lambda tz^{\rho-2}(\rho-2)^2)]$ tends to 0 with $t\to 0$. Hence, solution (6.4) tends to $C_1(x+s)^{-\gamma}$ at $t\to 0$. Next, for $u>1$, $E[u,v]\to 1/(u-1)$ at $v\to 0$, hence $E[(\gamma-1)/(\rho-2),1/(\lambda tz^{\rho-2}(\rho-2)^2)] >$const$>0$ at $t\to\infty$ if $2<\rho<1+\gamma$. Hence, the solution (6.4) is unbounded because this exponential integral is multiplied by the factor $(\lambda t)^{(\gamma-\rho+1)/(\rho-2)}$ that tends to infinity.
Q.E.D.

**Acknowledgements**
Authors thank Dr. E. Koonin for valuable discussions and help in preparation of the manuscript. The work of F. Berezovskaya was supported by NSF Grant #634156.